# ACCURACY VERSUS SPEED IN FLUCTUATION-ENHANCED SENSING

P. MAKRA [a], Z. TOPALIAN [b], C.G. GRANQVIST [b], L.B. KISH [c] and C. KWAN [d]

[a] *Department of Experimental Physics, University of Szeged, Dóm tér 9, Szeged, Hungary*
[b] *Department of Engineering Sciences, The Ångström Laboratory, Uppsala University, P.O. Box 534, SE-751 21 Uppsala, Sweden*
[c] *Department of Electrical and Computer Engineering, Texas A&M University, Mailstop 3128, College Station, TX 77841-3128, USA*
[d] *Signal Processing Inc, 13619 Valley Oak Circle, Rockville, MD 20850, USA*

Fluctuation-enhanced sensing comprises the analysis of the stochastic component of the sensor signal and the utilization of the microscopic dynamics of the interaction between the agent and the sensor. We study the relationship between the measurement time window and the statistical error of the measurement data in the simplest case, when the output is the mean-square value of the stochastic signal. This situation is relevant at any practical case when the time window is finite, for example, when a sampling of the output of a fluctuation-enhanced array takes place; or a single sensor's activation (temperature, etc) is stepped up; or a single sensor's output is monitored by sampling subsequently in different frequency windows. Our study provides a lower limit of the relative error versus data window size with different types of power density spectra: white noise, $1/f$ (flicker, pink) noise, and $1/f^2$ (red) noise spectra.

## 1. Introduction

Recently, a wide range of applications [1-34] has been introduced, where the frequency- [1-28], correlation- [29-30], or coincidence-analysis [31-34] of random [1-25,29-34] or randomness-reletad [26-28] fluctuations have been utilized as a source of information. The relevant generic fields are fluctuation-enhanced sensing [1-25]; vibration-induced conductance fluctuations [26-28]; noise-based logic and computation [29-32]; and unconditionally secure communications [33,34].

Fluctuation-enhanced sensing (FES) [1–25] means that the microscopic stochastic fluctuations superimposed on the sensor signal are extracted, processed and utilised as the sensory information carrier. FES has shown strongly increased sensitivity and selectivity with various film sensors and has the potential to replace a complex multisensor electronic nose by a single (or a few) sensor(s).

The actual information content of a FES signals is a subject of studies [2]. The present investigation is concerned with the situation when the accuracy of a simple statistics of FES-based frequency analysis is limited by the statistical inaccuracy due to limited sample size (limited time window length) and frequency bandwidth. Such situations can often occur, for example with sensor arrays wherein a single amplifier channel is sampling the sensors in a sequential mode; or when single-sensor selectivity parameters are varied in a range such as by modifying the activation of the sensor by sweeping the temperature, the electrical field, etc.

The accuracy versus the shape of the noise spectrum and its dependence on the sampling time window is of particular interest and is addressed in the analysis below.

**2. Goals**

In this Letter, we consider a fluctuation-enhanced sensing set-up for which the mean-square of the sensor noise provides the main sensory information. Since the power spectral density, a well-established and widely used tool in fluctuation-enhanced sensing, is basically the frequency-distribution of the mean-square, this simple model is relevant also for the accuracy of the sensory information obtained from the power density spectrum.

Whenever a fluctuating quantity is measured, taking longer samples increases the accuracy of the empirical statistics. Yet for practical applications a shorter sample length is desirable to speed up the detection and reduce the processing load. Below, we examine the means to find a reasonable compromise between these factors and also address the technical difficulties of such simulations.

**3. Modeling**

In our numerical simulations, we generated noise sequences of $N$ points, denoted by $\{x_j\}_{j=0}^{N-1}$, for which the sample length $N$ was varied in the course of the simulation. At a given sample length $N$, we generated $M$ independent samples and calculated the statistics of the mean-square. For each noise realisation $k$ ($0 \leq k < M$), we determined the mean of the squared values by

$$\mu_k(N) = E\{x_j^2\}_{j=0}^{N-1} = \frac{1}{N} \sum_{j=0}^{N-1} x_j^2 \ . \tag{1}$$

The more $\mu$ fluctuates the less reliable is the detection and agent recognition, so to describe the accuracy we can take the standard deviation of the mean-square obtained from the ensemble of $M$ realisations according to

$$D_\mu(N) = \sqrt{\frac{1}{M-1} \sum_{k=0}^{M-1} \left[\mu_k(N) - E_\mu(N)\right]^2} \ , \qquad (2)$$

where $E_\mu(N)$ is the ensemble average of the mean-square values $\mu_k(N)$, as expressed by

$$E_\mu(N) = \frac{1}{M} \sum_{k=0}^{M-1} \mu_k(N) \ . \qquad (3)$$

We chose the relative error of the mean-square, defined as the ratio of the standard deviation and the average of the mean-square, as the reliability parameter, ie,

$$\varepsilon(N) = \frac{D_\mu(N)}{E_\mu(N)} \ . \qquad (4)$$

Within this formalism, the aim of the simulation can be expressed as an exploration of the dependence of the relative error of the mean-square $\varepsilon(N)$ on the sample length $N$. The dependence of the mean of white noise on sample length is a basic formula of statistics ($D_{\text{mean}} = \sigma/\sqrt{N}$, where $\sigma$ is the standard deviation of the white noise), but in our case a numerical simulation is warranted as sensor noise is rarely white, and for coloured noise this relationship does not hold.

We carried out our numerical simulations and, as mentioned above, we examined three different types of noise: Gaussian white noise, $1/f$ noise and Brownian motion or $1/f^2$ noise.

Coloured noise proved somewhat problematic. Originally, we used a simple spectral filtering algorithm [35] to generate $1/f^\kappa$ power law noise, but analysis of the preliminary data revealed that noise obtained by filtering Gaussian white noise will always reflect the statistics of the mean of the original white noise. The original filtering algorithm simply took a white noise sequence $\{w_j\}_{j=0}^{N-1}$, Fourier transformed it, and scaled the discrete Fourier components $W_k$ in inverse proportion to the index representing the frequency raised to half the desired power exponent $\kappa$, ie,

$$X_k = W_k / k^{\frac{\kappa}{2}} \qquad (1 \le k \le \frac{N}{2}),$$

whilst the second half of the spectrum was obtained from the symmetry relation that holds for Fourier transforms of real signals:

$$X_k = \overline{X_{N-k}} \qquad (\frac{N}{2} < k < N),$$

where the overline denotes the complex conjugate. In this scheme, the $0^{th}$ component of the Fourier transform clearly represents a singularity, so we either keep the $W_0$ component of the original white noise or assign a constant value to it. Unfortunately, this $0^{th}$ component is proportional (or equal, depending on the choice of normalisation in the Discrete Fourier Transform) to the mean value $\mu$ of the noise:

$$X_0 = \sum_{j=0}^{N-1} x_j e^{i \cdot j \cdot 0 \frac{2\pi}{N}} = \sum_{j=0}^{N-1} x_j = N \cdot \mu.$$

This means, if we keep the original $0^{th}$ component, we will have a noise whose mean will be exactly the same as the mean of the original white noise and consequently will have the same statistics and dependence on the length of the sample, whilst if we assign a constant value to the $0^{th}$ component, the mean of the resultant colored noise will be exactly defined and will not fluctuate at all, thus showing no dependence whatsoever on sample length. As a result, the spectral filtering algorithm is utterly unfit for the purposes of investigating the statistics of the mean.

For this reason we had to take a different route in coloured noise generation. To represent $1/f^2$ noise, we took a simple random walk, defined by

$$b_i = b_{i-1} + u_i \qquad (0 < i < N), \tag{5}$$

where $u_i$ is a random number of a uniform distribution between -0.5 and 0.5. To generate $1/f$ noise, we adopted Phil Burk's optimised version [36] of the Voss method, which is based on adding up uniform random numbers evaluated in octave time intervals [37].

The parameters of the simulation were the following: the shortest sample length was 10, which we increased in steps of 10 up to a maximum sample length of 1000. For each sample length, we generated an ensemble of 100000 noise realisations from which we determined the error and the average of the mean-square value of the samples depending on the sample length $N$.

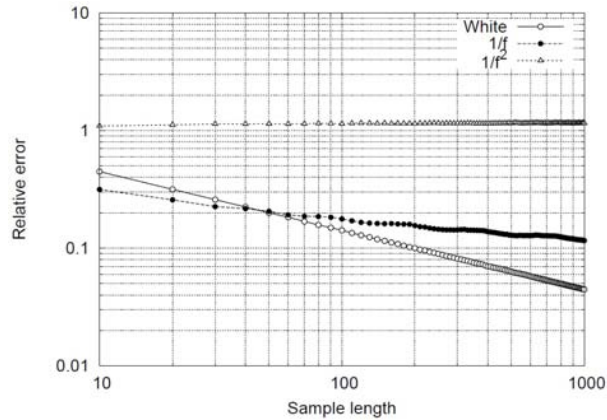

**Figure 1**. Relative error of noise detection as a function of the length of noise samples.

## 3 Results

Fig. 1 sums up our results for the three types of noise. We can see that the relative error of the mean-square decreases steadily in the case of white noise, whilst for $1/f^{\,2}$ noise taking a longer sample has no effect whatsoever on the relative error of the mean-square. The $1/f$ noise is somewhere in between: the relative error does decrease with sample length, yet less steeply than for white noise.

As sensor noise is most often coloured [1,3-9], its behaviour is expected to fall in the range between the extremes represented by white noise and Brownian motion, but here we can only choose $1/f$ noise as a representative. We can see that the scope for accuracy improvement is limited: at the cost of reducing the speed by a factor of 100, the relative error only decreases by a factor of 2.7 in the case of $1/f$ noise, whereas for white noise this factor is 10.

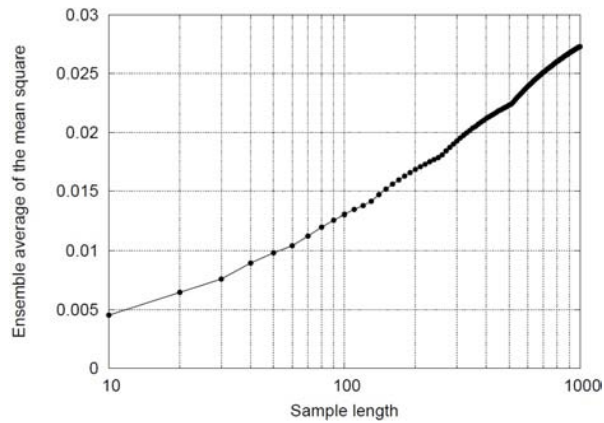

**Figure 2.** Average value of the mean-square of the 1/*f* noise as a function of the sample length.

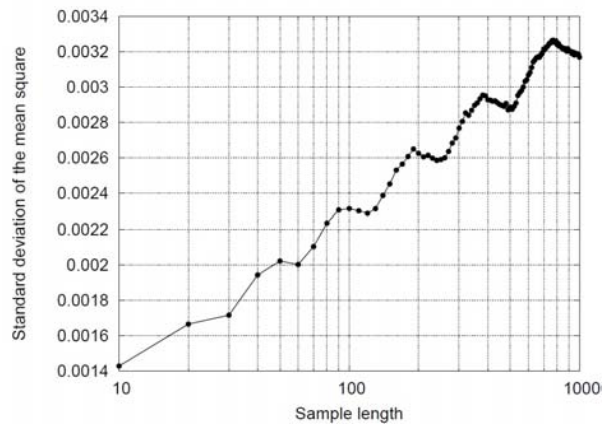

**Figure 3.** Error of the mean-square as a function of the sample length for the simulated 1/*f* noise.

The decrease of the relative error in the case of 1/*f* noise can result from both a decrease in the error of the mean-square $D_\mu$ and an increase in the mean-square value $E_\mu$ itself. The latter is also predicted by theory [38]: the mean-square can also be calculated from the power spectral density $S(f)$ by

$$E_\mu = \int_{f_{\min}}^{f_{\max}} S(f)\mathrm{d}f = \int_{f_{\min}}^{f_{\max}} \frac{C}{f}\mathrm{d}f = C\ln\left(\frac{f_{\max}}{f_{\min}}\right), \qquad (6)$$

where $C$ is a constant while $f_{\min}$ and $f_{\max}$ denote the minimum and maximum frequencies, respectively. These latter entities are determined by the sample length $N$ in a discrete realisation. Let us denote the sampling interval by $\Delta t$. In this case, the minimum frequency is

$$f_{\min} = \frac{1}{T} = \frac{1}{N\Delta t}, \qquad (7)$$

whilst the maximum frequency is the Nyquist frequency given by

$$f_{\max} = \frac{f_s}{2} = \frac{1}{2\Delta t}, \qquad (8)$$

where $T$ denotes the measurement time and $f_s = \frac{1}{\Delta t}$ is the sampling frequency. From these expressions it follows [38] that the mean-square value of the white noise is

$$E_\mu(N) = C\ln\left(\frac{N}{2}\right). \qquad (9)$$

Fig. 2 shows that the mean-square value of $1/f$ noise indeed increases with sample length, although for small sample lengths the simulation results deviate from the logarithmic dependence predicted by theory. The deviation of the plot Fig. 2 from a straight line indicates that the Voss method produces only approximate $1/f$ noise when the time window is limited. Furthermore, Fig. 3 illustrates that the "octave time intervals" sampling of the Voss method has an impact on the error of the mean-square, where some periodic platoes appear. However, it is important to note that typical FES signals are also deviate from 1/f noise and our study is only tentative. The results furthermore indicate that the mechanism of relative error reduction with $1/f$ like FES is different from that of white noise: it is not the decrease of the error of the mean that causes the decrease of the relative error but the increase of the mean-square itself.

## 4 Conclusion

We carried out numerical simulations to investigate how the accuracy of noise measurements depends on the length of the noise sample for different noise types: Gaussian white, $1/f$ and Brownian noise. Our results indicate that for $1/f$ noise, which can be taken as a model for sensor noises, increasing the sample length yields only a moderate improvement in reliability. Moreover, we have shown that this improvement does not imply that the mean-square fluctuates to a lesser extent if we take longer samples; in fact, the standard deviation of the mean-square increases with sample length, but at the same time the value of the mean-square itself is also a function of sample length and it increases at a higher rate.

**Acknowledgements**

Financial support was received from the European Research Council under the European Community's Seventh Framework Program (FP7/2007-2013)/ERC Grant Agreement 267234 ("GRINDOOR") and from the National Institute of Standards under contract SB134111SE0884.